\documentclass[aps, twocolumn,superscriptaddress]{revtex4-1}

\usepackage{lipsum}
\usepackage{braket}
\usepackage{amsmath}
\usepackage{amssymb}
\usepackage{hyperref}
\usepackage{siunitx}
\usepackage[normalem]{ulem}
\usepackage{graphicx}
\usepackage{colortbl}
\usepackage{xcolor}

\begin{document}
\title{Hessian-based optimization of constrained quantum control}
\author{Mogens Dalgaard}
\affiliation{Department of Physics and Astronomy, Aarhus University, Ny Munkegade 120, 8000 Arhus C, Denmark}
\date{July 2019}
\author{Felix Motzoi}
\affiliation{Forschungszentrum J\"ulich, Institute of Quantum Control (PGI-8), D-52425 J\"ulich, Germany}
\author{Jesper Hasseriis Mohr Jensen}
\affiliation{Department of Physics and Astronomy, Aarhus University, Ny Munkegade 120, 8000 Arhus C, Denmark}
\date{July 2019}
\author{Jacob Sherson}
\email{sherson@phys.au.dk}
\affiliation{Department of Physics and Astronomy, Aarhus University, Ny Munkegade 120, 8000 Arhus C, Denmark}
\date{May 2019}

\begin{abstract}
    Efficient optimization of quantum systems is a necessity for reaching fault tolerant thresholds. A standard tool for optimizing simulated quantum dynamics is the gradient-based \textsc{grape} algorithm, which has been successfully applied in a wide range of different branches of quantum physics. In this work, we derive and implement exact $2^{\mathrm{nd}}$ order analytical derivatives of the coherent dynamics and find improvements compared to the standard of optimizing with the approximate  $2^{\mathrm{nd}}$ order \textsc{bfgs}. We demonstrate performance improvements for both the best and average errors of constrained unitary gate synthesis on a circuit-\textsc{qed}
    system over a broad range of different gate durations.
\end{abstract}
\maketitle
\section{Introduction}

Nowadays, there exists a wide range of proposed technologies that utilize the potential of quantum mechanics in order to achieve improvements over their classical counterparts. These include quantum variational eigensolvers \cite{peruzzo2014variational,kandala2017hardware}, annealers \cite{johnson2011quantum}, simulators \cite{houck2012chip,bloch2012quantum}, Boltzman machines \cite{amin2018quantum}, and perhaps most promising, the quantum computer \cite{nielsen2002quantum}. We may reach a point in time where the majority of these quantum-based technologies outperform their classical counterparts.

Reaching this \textit{quantum advantage} requires, among others, substantial improvements in our ability to control the underlying quantum systems. On the theoretical side, quantum optimal control theory addresses this issue \cite{werschnik2007quantum, brif2010control, glaser2015training}. Here optimization methods with respect to a chopped random basis (\textsc{crab}) \cite{doria2011optimal,caneva2011chopped} and individual pulse amplitudes (Krotov) \cite{palao2002quantum,palao2003optimal} have been successful especially with respect to unconstrained quantum optimization, where trapping has been shown to rarely occur \cite{rabitz2004quantum, rach2015dressing}. 

One of the most used, and widely successful, algorithms within quantum optimal control theory is the gradient ascent pulse engineering (\textsc{grape}) algorithm \cite{khaneja2005optimal,de2011second}. The original \textsc{grape} algorithm used first-order approximated gradients in combination with steepest descent \cite{khaneja2005optimal}. Later, significant improvements were obtained when the analytical gradients were calculated and combined with Hessian approximation methods such as \textsc{bfgs} \cite{de2011second,machnes2011comparing, fletcher2013practical}. \textsc{grape} has been widely successful providing its use in Nuclear Magnetic Resonance \cite{tovsner2006effective, zhang2011experimental,kobzar2012exploring,rao2014efficient,sogaard2019optimization}, superconducting qubit circuits \cite{motzoi2009leakage,groszkowski2011tunable,egger2013optimized,kirchhoff2018optimized,abdelhafez2020universal}, spin chains \cite{paz2009perfect, wang2010robust,nimbalkar2012multiple,ashhab2015quantum}, Nitrogen vacancy centers \cite{said2009robust,dolde2014high}, and ultra cold atoms \cite{khani2012high,jensen2019time}. It has also become a standard integrated tool in many numerical packages aimed at quantum physicists \cite{tovsner2009optimal, machnes2011comparing, johansson2013qutip, sorensen2019qengine}. Further, one could mention the many extensions of \textsc{grape}, which treat filtering \cite{motzoi2011optimal}, robustness \cite{khaneja2005optimal,ge2020robust}, chopped random basis \cite{sorensen2018quantum}, experiment design \cite{goerz2017charting}, and open quantum systems \cite{schulte2011optimal}. 

In addition, there exists many hybrid algorithms that combine \textsc{grape} with e.g.~sequential updates \cite{machnes2011comparing}, and global optimization algorithms \cite{sorensen2018approaching,zahedinejad2014evolutionary}. Machine-learning based control \cite{palittapongarnpim2017learning, bukov2018reinforcement, fitzek2019deep, niu2019universal,an2019deep, yao2020policy, dalgaard2019global} could also improve the solution exploration of local quantum optimal control methods such as \textsc{grape}, where initial steps towards a hybrid algorithm have already been taken in Ref. \cite{dalgaard2019global}. 

In this paper we show how \textsc{grape} can further be enhanced to significantly increase its ability to reach high fidelity solutions, by incorporating the exact Hessian in the numerical optimization. We present here an efficient calculation of the Hessian and apply both gradient and Hessian-based optimization to unitary gate synthesis for superconducting circuits. A previous calculation of the Hessian made use of the auxiliary matrix method \cite{goodwin2015auxiliary,goodwin2016modified}. This approach requires the exponentiation of block matrices having three times the size of the Hilbert space, which leads to an unfavorable scaling both with respect to the Hilbert space size and the number of controls. In this work, we present an new derivation that only requires exponentiation of matrices with the same size as the Hilbert space. In doing so, we show that every component needed to evaluate the gradient can be recycled to evaluate the Hessian and thus that calculating an element of the Hessian is not much more expensive than the gradient. These results enable us to demonstrate improvements with Hessian-based \textsc{grape} in equal wall-time simulations without the explicit need for any code parallelization. In addition to enhancing optimization, the Hessian can also be used to characterize the quantum optimal control landscape, which in Ref. \cite{hocker2014characterization} was done in the presence of noise. 

Besides calculating the Hessian, we further seek to benchmark analytical gradient- and Hessian-based numerical optimization within constrained physical settings. Physically realizable pulses are always constrained in amplitude due to either limitations on experimental equipment, but more often imposed to avoid undesirable processes such as leakage, ionization, heating, decoherence, and break down of theoretical models. For many quantum control problems, optimal solutions contain segments that lie on the boundary of the admissible region (see e.g.~\cite{hegerfeldt2013driving, lin2020time}). Under certain assumptions, this has even been proven to more generally occur \cite{zhdanov2017structure}. Therefore, it is important how the optimization algorithms handle constraints.

\begin{figure}
    \centering
    \includegraphics{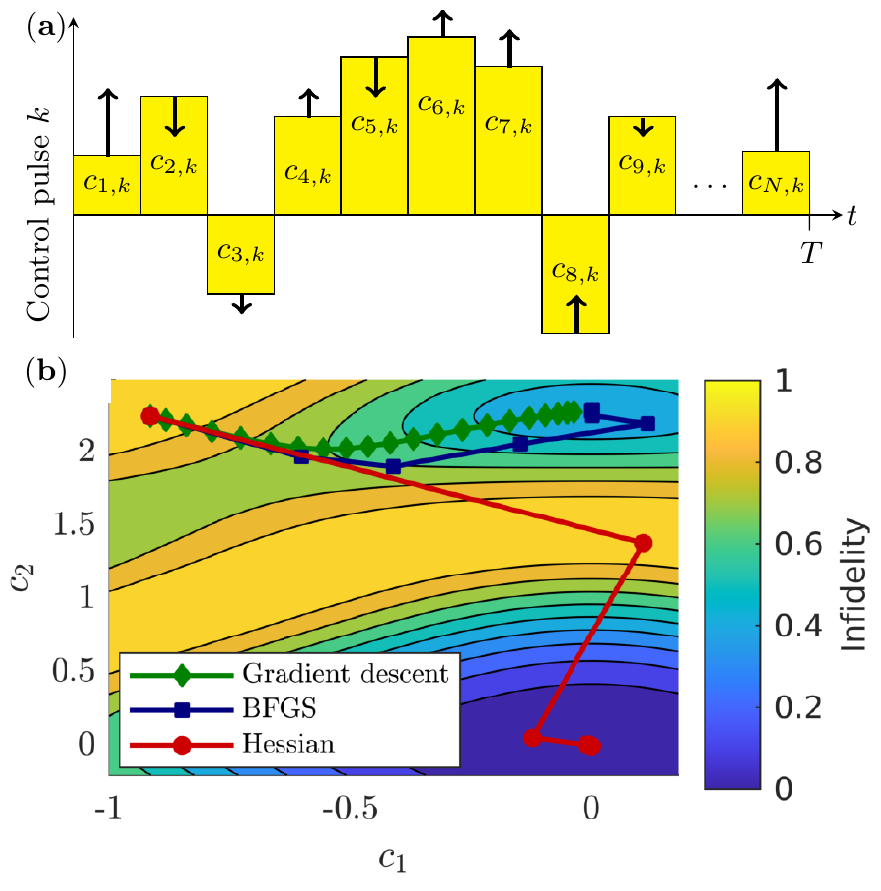}
    \caption{(Color online) $\mathbf{(a)}$ An example of a piecewise constant control pulse. Here individual updates are depicted as arrows. $\mathbf{(b)}$. An example of how Interior-Point with either \textsc{bfgs} or the Hessian searches differently. In addition we also compare to a simple gradient-descent algorithm. We elaborate on this figure in Appendix \ref{sec:two_control_example}.}
    \label{fig:pulse_ex_and_grad_v_Hess}
\end{figure}

For the work presented here, we demonstrate our methods by synthesizing the unitary dynamics of a superconducting transmon circuit \cite{koch2007charge}. The circuit consist of two fixed frequency transmon qubits dispersively coupled through a microwave resonator \cite{ rigetti2010fully, chow2011simple, goerz2017charting}. This setup could be used as a subpart of many of the aforementioned quantum technologies.

\section{Optimal control of unitary gates} \label{sec:optimal_control_of_unitary}

Our objective is to generate a target unitary $V$ using a set of piecewise constant pulses. However, it should be mentioned that the calculations given here are applicable to many other control problems as well. These include state-to-state transfer for pure quantum states \cite{jensen2020exact} (see also Appendix \ref{sec:state_to_state}) and density matrices in closed quantum systems, as well as state-to-state transfer of density matrices in open quantum systems \cite{khaneja2005optimal,machnes2011comparing}.

The pulses consist of $N$ constant-amplitude segments and have a total duration $T$. We assume access to a set of $M$ control Hamiltonians $\{H_k \}_{k=1}^M$ such that a bilinear Hamiltonian $H(t_j) = H_0 + \sum_{k=1}^M c_{j,k} H_k$ governs the system dynamics at time step $j$. Here $H_0$ and $c_{j,k}$ denotes the drift Hamiltonian and controls respectively. We have depicted an example of a control pulse in Fig. \ref{fig:pulse_ex_and_grad_v_Hess}(a). The system starts from an initial unitary $U_0$, which is typically chosen to be the identity, and then evolves through unitary evolution $U_j = \exp(-iH(t_j) \Delta t)$ where $\Delta t = T/N$ and $\hbar = 1$.

Replicating the target unitary $V$, up to a global phase, is achieved by maximizing the fidelity 

\begin{equation}
    \mathcal{F} = \Big|\frac{1}{\text{dim}} \text{Tr}[\mathbb{P}U\mathbb{P} V^{\dagger}]\Big|^2,
    \label{eq:Fidelity}
\end{equation}
where $U = U_N U_{N-1}\ldots U_0$ denotes the final unitary and $\mathbb{P}$ a projection into the subspace of interest, whose dimension is denoted $\text{dim}$. This approach exploits that quantum propagation can always be decomposed smoothly into shorter, differential time segments. Equivalently, we seek to minimize the infidelity $J = 1-\mathcal{F}$, which will serve as our cost function. We express a control vector as $\mathbf{c} = (c_{1,1}, c_{2,1}, \ldots, c_{N,1},c_{1,2}, \ldots c_{N,M}  )$, where the first index denotes the discretized time and the second denotes different controls. Starting from an initial guess $\mathbf{c}_0$, we seek to make incremental updates such that $J(\mathbf{c}_{n+1}) < J(\mathbf{c}_{n})$, with $n$ denoting the iteration number. The incremental update is on the form 
\begin{align}
    \mathbf{c}_{n+1} = \mathbf{c}_{n} + \alpha_n \mathbf{p}_{n}.
\end{align}
Here $\mathbf{p}_{n}$ defines a search direction, while $\alpha_n$ defines a step size typically found through a line-search. Fig.~\ref{fig:pulse_ex_and_grad_v_Hess}(a) illustrates an example of an incremental update depicted with arrows. The initial seed could e.g. stem from an analytical ansatz or based on a random guess to explore the optimization landscape. The latter approach is termed multistarting \cite{ugray2007scatter}.

The search direction taken by \textsc{grape} was originally proposed to be steepest descent, $\mathbf{p}_n = - \nabla J(\mathbf{c}_n)$, where the gradient was approximated to first order in $\Delta t$ \cite{khaneja2005optimal}. Steepest descent only uses information about the first derivative and thus suffers from having no information about the curvature of the optimization landscape. Hence, it is often superior to use the $2^{\mathrm{nd}}$ order derivatives, i.e.~Hessian matrix

\begin{align}
    \mathbf{H} =
    \begin{bmatrix}
    \frac{\partial^2 J}{\partial c_{1,1}^2 } 
    & \frac{\partial^2 J}{\partial c_{1,1} \partial c_{2,1}} 
    &\ldots
    &\frac{\partial^2 J}{\partial c_{1,1} \partial c_{N,M} }\\
    \frac{\partial^2 J}{\partial c_{2,1} \partial c_{1,1}}
    &\frac{\partial^2 J}{\partial c_{2,1}^2 } 
    & \ldots  
    & \frac{\partial^2 J}{\partial c_{2,1} \partial c_{N,M}}\\
    \vdots & \vdots & \ddots & \vdots\\
    \frac{\partial^2 J}{\partial c_{N,M} \partial c_{1,1}}
    & \frac{\partial^2 J}{\partial c_{M,N} \partial c_{2,1}}
    & \ldots 
    & \frac{\partial^2 J}{\partial c_{N,M}^2 }
    \end{bmatrix}.
\end{align}
If one has access to the Hessian, a standard search direction is given by Newton’s method $\mathbf{p}_n = - [\mathbf{H}(\mathbf{c_n})]^{-1} \nabla J(\mathbf{c}_n)$. If the $2^{\mathrm{nd}}$ order derivatives are not available, one can approximate the Hessian $\mathbf{B} \approx \mathbf{H}$ via the \textsc{bfgs} Hessian-approximation scheme \cite{fletcher2013practical} that gradually builds $\mathbf{B}$ using only the gradient. The iterative \textsc{bfgs} update is of the form

\begin{align}
    \mathbf{B}_{k+1} = \mathbf{B}_{k} +  
    \frac{\mathbf{y}_k \mathbf{y}_k^T}{\mathbf{y}_k^T \mathbf{s}_k}
    - \frac{\mathbf{B}_{k} \mathbf{s}_k \mathbf{s}_k^T \mathbf{B}_{k}^T }
    {\mathbf{s}_k^T \mathbf{B}_{k} \mathbf{s}_k},
\end{align}
where $\mathbf{y}_k = \nabla f(\mathbf{x}_{k+1}) - \nabla f(\mathbf{x}_{k})$ and $\mathbf{s}_k = \mathbf{x}_{k+1} - \mathbf{x}_{k}$. The derivation and use of an exact gradient enabled the precision needed for \textsc{bfgs} to outclass its first-order counterpart \cite{de2011second}. It is now the standard within quantum optimal control theory \cite{machnes2011comparing, johansson2013qutip, sorensen2019qengine}.

Since physical realizable pulses are always constrained in amplitude, we optimize via a constrained optimization algorithm that uses either the actual Hessian or the gradient-based \textsc{bfgs} (see Appendix \ref{sec:different_optim}). In Fig.~\ref{fig:pulse_ex_and_grad_v_Hess}(b) we illustrate the difference between these methods (Hessian-based and gradient-only) for a simple two-dimensional optimization problem (see Appendix \ref{sec:two_control_example} for a technical explanation). The problem has a global minimum at $(c_1,c_2) = (0,0)$. In addition, there is also a local one above. We start the optimization near the two solutions and plot how each method moves through the optimization landscape. The two gradient-based approaches converge toward the local solution, while the Hessian-based method manages to avoid this solution and converge towards the global optimum. A possible explanation for this behavior is that Hessian-based optimization generally has more information about the landscape curvature. This can allow it to avoid suboptimal nearby solutions as the one depicted in Fig.~\ref{fig:pulse_ex_and_grad_v_Hess}(b). As we demonstrate later, the two different approaches (gradient-only and Hessian based optimization) often find different solutions although starting from the same initial pulse. However, we would like to stress that the Hessian is not guaranteed to find the best solution of the two.

Moreover, Fig.~\ref{fig:pulse_ex_and_grad_v_Hess}(b) also illustrates how Hessian-based optimization generally requires fewer iterations to converge \cite{nocedal2006numerical}. If the cost function is sufficiently simple near an optimal solution $\mathbf{x}^*$ a nearby initial guess (seed) $\mathbf{x}_0$ will converge as $||\mathbf{x}_{k+1}-\mathbf{x}^*|| \leq C||\mathbf{x}_{k}-\mathbf{x}^*||^q$ where $C\geq 0$ for the optimization methods discussed here, with Hessian being quadratic ($q = 2$) and \textsc{bfgs} being superlinear ($1 < q < 2$) \cite{nocedal2006numerical}.

In the following we derive exact analytical expressions for both the gradient and Hessian of Eq.~(\ref{eq:Fidelity}). The gradient has previously been derived, see e.g.~Refs. \cite{de2011second,machnes2011comparing}, but reproduced here for completeness. 

\subsection{Gradient}
The first step is to express the derivative of the fidelity, Eq. (\ref{eq:Fidelity}), in terms of the derivative of the unitary 

\begin{align}   
    \frac{\partial \mathcal{F}}{\partial c_{j,k}} 
    &=  
    \frac{2}{\text{dim}^2}
    \text{Re}
    \Bigg[
    \text{Tr} \Big[\mathbb{P}\frac{\partial U}{\partial c_{j,k}}\mathbb{P} V^{\dagger} \Big]
    \text{Tr}\Big[V \mathbb{P}U^{\dagger}\mathbb{P}\Big]
    \Bigg].
    \label{eq:derive_fidelity}
\end{align}
It is only the unitary at the $j$th time-step, $U_j$, that depends on $c_{j,k}$ and hence the derivative is
\begin{equation}
    \frac{\partial U}{\partial c_{j,k}} = U_N \ldots \frac{\partial U_j}{\partial c_{j,k}} 
    \ldots U_1 U_0 
    = U^L_{j+1} \frac{\partial U_j}{\partial c_{j,k}} U^R_{j-1},
    \label{eq:derive_final_unitary}
\end{equation}
where we have defined the left and right unitaries as $U^L_{j} = U_N U_{N-1} \ldots U_j$ and $U^R_j = U_j U_{j-1} \ldots U_0$, respectively. The left and right unitaries must be calculated for each time step, but this can be done efficiently since $U^L_{j} = U^L_{j-1}U_j$ and $U^R_j = U_j U^R_{j-1}$. Thus, the gradient calculations scales as $\mathcal{O}(N)$ in terms of the number of matrix multiplications rather than the intuitive $\mathcal{O}(N^2)$.

In order to evaluate the derivative $\frac{\partial U_j}{\partial c_{j,k}}$ we use the following expression for a general $\eta$-dependent matrix $\chi(\eta)$ \cite{wilcox1967exponential}

\begin{equation}
     \frac{d}{d \eta} e^{\chi(\eta)} = \int_0^{1} e^{\alpha \chi(\eta)} \frac{d \chi(\eta)}{d \eta} e^{(1-\alpha) \chi(\eta)} d \alpha.
    \label{eq:matrix_derivative}
\end{equation}
There are many known ways to approximate this integral (see Appendix \ref{sec:expansion}), especially relevant when state transfer or sparse matrices are optimized (see Appendix \ref{sec:state_to_state}). Here, however, we consider an exact method, which is to solve it explicitly in the eigenbasis $\{ \ket{n} \}$ of the Hamiltonian $H(t_j)$. A direct calculation reveals

\begin{equation}
    \Braket{m| \frac{\partial U_j}{\partial c_{j,k}} |n} = \Braket{m|H_k|n} I(m,n).
    \label{eq:GRAPE_gradient}
\end{equation}
Here we have defined
\begin{align}
    I(m,n) =
    \begin{cases}
    -i \Delta t e^{-i E_m \Delta t }, &\textit{  if } E_m=E_n\\
    \frac{e^{-i E_m \Delta t } - e^{-i E_n \Delta t }}{E_m - E_n}, &\textit{  if } E_m \neq E_n,  
    \end{cases}
    \label{eq:Integral_gradient}
\end{align}
where $E_n$ denotes the eigenenergy of $\ket{n}$. This method requires diagonalization of the Hamiltonian at each instance of time i.e.~$H(t_j) = R^{\dagger} D R$, with $R$ being transformation matrix whose columns are the eigenvectors of $H(t_j)$ and $D$ containing the eigenenergies $D = \text{Diag}(E_1,E_2,\ldots)$. One can use the diagonalization efficiently to first evaluate the matrix exponential $e^{-iH(t_j) \Delta t} = R^{\dagger} \text{Diag}(-i E_1 \Delta t, -i E_2 \Delta t,\ldots) R$, and subsequently use it to switch from one basis to another. If we define the matrix $I$ with $I_{m,n} = I(m,n)$, Eq.~(\ref{eq:GRAPE_gradient}) can be written in the original basis as

\begin{align}
    \frac{\partial U_j}{\partial c_{j,k}} = R\left((R^{\dagger} H_k R) \odot I \right) R^{\dagger},
    \label{eq:GRAPE_gradient_originial_basis}
\end{align}
where $\odot$ denotes the Hadamard (element-wise) product. For multiple controls $\{ H_k \}_{k=1}^M$ each derivative $\{ \partial U_j/\partial c_{j,k} \}_{k=1}^M$ may be evaluated efficiently since $R, R^{\dagger}$, and $I$ need only be calculated once. Note one can also exploit the symmetry $I_{m,n} = I_{n,m}$ when evaluating $I$.

In the case of larger Hilbert spaces, the exact matrix diagonalization becomes intractable. In this case, there exist other methods in the literature that may be more applicable while still permitting analytical differentiation \cite{floether2012robust, goodwin2018advanced, jensen2020exact} and gradient and Hessian-based optimization.

\subsection{Hessian}

\begin{figure*}
    \centering
    \includegraphics{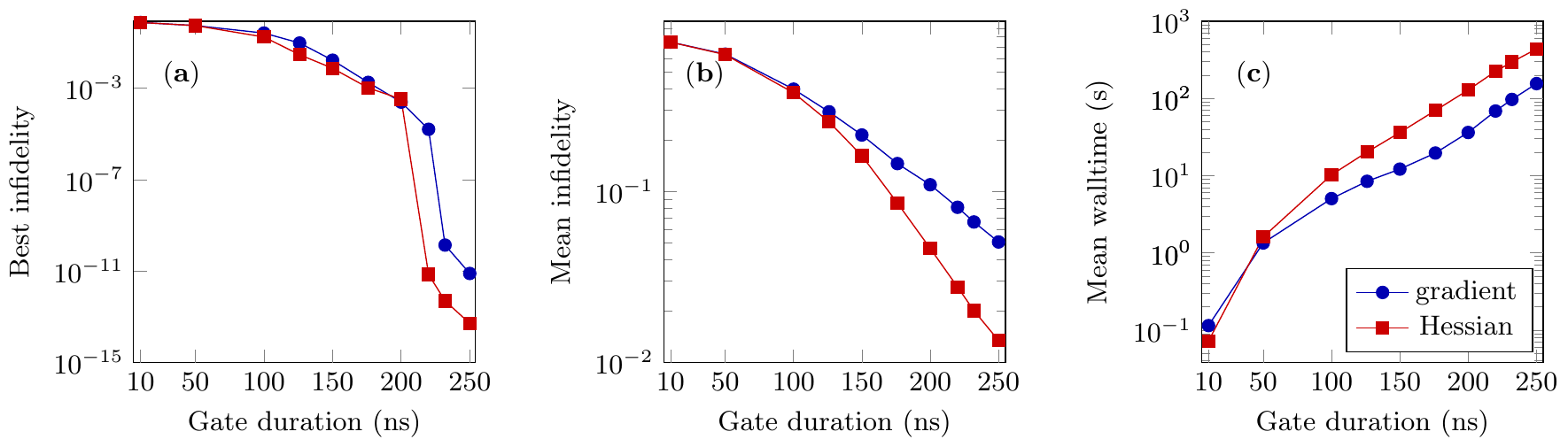}
    \caption{The results of \textsc{grape} optimizations using Interior-Point with either the Hessian or the gradient-based Hessian-approximation scheme \textsc{bfgs} (denoted gradient) for the same 5000 seeds which we draw uniformly at random for each gate duration. The figure depicts $\mathbf{(a)}$ the best infidelity, $\mathbf{(b)}$ the mean infidelity per seed, and $\mathbf{(c)}$ the mean wall time consumption per seed.}
    \label{fig:grad_Hess_comparison_different_gate_times}
\end{figure*}

We start by evaluating the derivative of Eq. (\ref{eq:derive_fidelity}), which reveals
\begin{align} \nonumber
    \frac{\partial^2 \mathcal{F}}{\partial c_{i,k'} \partial c_{j,k}} 
    &= \frac{2}{\text{dim}^2} \text{Re}
    \Bigg[
    \text{Tr} 
    \Big[ \mathbb{P}
    \frac{\partial U^2}{\partial c_{i,k'} \partial c_{j,k}} \mathbb{P} V^{\dagger}
    \Big] 
    \text{Tr} 
    \Big[ 
    V \mathbb{P} U^{\dagger} \mathbb{P}
    \Big]  \\
    &+
    \text{Tr} 
    \Big[ \mathbb{P}
    \frac{\partial U}{\partial c_{j,k}} \mathbb{P} V^{\dagger}
    \Big]   
    \text{Tr} 
    \Big[ \mathbb{P}
    \frac{\partial U}{\partial c_{i,k'}} \mathbb{P} V^{\dagger}
    \Big]^*
    \Bigg].
\end{align}
The first order derivatives of the unitary have already been calculated when evaluating the gradient. What remains is hence to calculate the second derivatives of the unitary. Here we distinguish between two cases: when the derivatives are with respect to different time steps ($i \neq j$) or the same time step ($i = j$). For different time steps ($j > i$) the $2^{\mathrm{nd}}$ order derivatives become

\begin{align} 
     \frac{\partial^2 U}{\partial c_{i,k'} \partial c_{j,k} } 
    =U^L_{j+1} \frac{\partial U_j}{\partial c_{j,k}} U^R_{j-1} (U^R_{i})^\dagger
    \frac{\partial U_i}{\partial c_{i,k'}} U^R_{i-1}.
\label{eq:second_derive_U_different_steps}
\end{align}
Here we have expressed the middle products of unitaries $U_{j-1}\ldots  U_{i+1}$ in terms of the right unitaries we defined when calculating the gradient. 
Evaluating the $2^{\mathrm{nd}}$ order derivative of the unitary with respect to the same time step is similar to Eq. (\ref{eq:derive_final_unitary})

\begin{equation}
    \frac{\partial^2 U}{\partial c_{j,k'} \partial c_{j,k}} = 
    U^L_{j+1} \frac{\partial^2 U_j}{\partial c_{j,k'} \partial c_{j,k}} U^R_{j-1}.
    \label{eq:second_derive_U_same_steps}
\end{equation}
Eq. (\ref{eq:second_derive_U_different_steps}) and (\ref{eq:second_derive_U_same_steps}) imply that the Hessian may be evaluated efficiently by recycling already calculated quantities. All that is left is to calculate the $2^{\mathrm{nd}}$ order derivative of $U_j$. This is done by differentiating Eq. (\ref{eq:matrix_derivative})

\begin{align} \nonumber
    \frac{\partial^2 U_j}{\partial c_{j,k'} c_{j,k}}
    &= -i \Delta t \int_0^{1} 
    \left(
    \frac{\partial }{\partial c_{j,k'}}
    e^{\alpha X} 
    \right)
    H_k e^{(1-\alpha) X} d \alpha\\
    &
     -i \Delta t \int_0^{1} e^{\alpha X} H_k 
    \left(
    \frac{\partial }{\partial c_{j,k'}}
    e^{(1-\alpha) X} 
    \right)
    d \alpha,\end{align}
where  we have introduced the short-hand notation $X = -iH\Delta t$. In the last integral above, we make the substitution $1-\alpha \rightarrow \alpha$ and then insert Eq.~(\ref{eq:matrix_derivative}) to obtain

\begin{align} \nonumber
    &\frac{\partial^2 U_j}{\partial c_{j,k'} c_{j,k}}
    =\\  \nonumber
    &(-i \Delta t)^2 
    \Big(
    \int_0^1 d\alpha \alpha 
    \int_0^1 d\beta 
    e^{\beta \alpha X} 
    H_{k'}
    e^{(1-\beta) \alpha X}
    H_k
    e^{(1-\alpha)  X} \\ 
    &+
    \int_0^1 d\alpha \alpha 
    \int_0^1 d\beta 
    e^{( 1 - \alpha) X} 
    H_{k}
    e^{\beta \alpha X}
    H_{k'}
    e^{(1-\beta) \alpha X}
    \Big).
\end{align}

Similar to before, there are different ways to approximate this integral depending on how the exponential is propagated (see Appendices \ref{sec:state_to_state} and \ref{sec:expansion}). We derive here the exact solution by evaluating the elements of the eigenbasis of $H$. We insert the identity $\mathbf{1} = \sum_{n'} \ket{n'}\bra{n'}$
to evaluate the middle part i.e. between the two Hamiltonians $H_k$ and $H_{k'}$ in the above expression. This reveals

\begin{align}\nonumber
    &\Braket{m|
    \frac{\partial^2 U_j}{\partial c_{j,k'} c_{j,k}}
    |n}
    = (-i \Delta t)^2
    \sum_{n'}
    \Big(
    \\ \nonumber
    &
    H_{k'}^{(m,n')} H_{k}^{(n',n)}
    \int_0^1 d\alpha \alpha 
    \int_0^1 d\beta 
    e^{\beta \alpha \lambda_m}
    e^{(1-\beta) \alpha \lambda_{n'}}
        e^{(1-\alpha)  \lambda_n} \\ \nonumber
    &+\\\nonumber
    &H_{k}^{(m,n')} H_{k'}^{(n',n)}
    \int_0^1 d\alpha \alpha 
    \int_0^1 d\beta 
    e^{( 1 - \alpha) \lambda_m}
    e^{\beta \alpha \lambda_{n'}}
    e^{(1-\beta) \alpha \lambda_n}
    \Big),\\ 
    \label{eq:double_integrals}
\end{align}
where we have defined $H_k^{(m,n)} = \braket{m|H_k|n}$ and $\lambda_n = -iE_n\Delta t$. The two double integrals in the above expression turn out to be equivalent, which we calculate in Appendix \ref{sec:Hess}. This allows us to write
\begin{align}\nonumber
    &\Braket{m |
    \frac{\partial^2 U_j}{\partial c_{j,k'} c_{j,k}}
    |n}
    = 
    \\
    &\sum_{n'} \Big(
    H_{k'}^{(m,n')} H_{k}^{(n',n)} 
    +
    H_{k}^{(m,n')} H_{k'}^{(n',n)}
    \Big)\mathcal{I}(n,n',m).
    \label{eq:U_final_expression_of_elements}
\end{align}
Here $\mathcal{I}(n,n',m)$ is given by Eq. (\ref{eq:hessian_integral}). 
The intuitive scaling for calculating the analytical Hessian would be $\mathcal{O}(N^3)$, since one would have to calculate the left, middle, and right sequence of unitaries for each pair of time steps $(i,j)$. However, having the left and right unitaries in advance reduces this to $\mathcal{O}(N^2)$ via Eq.~(\ref{eq:second_derive_U_different_steps}). Note that this is also true for other methods for evaluating the time-ordered integral (Appendix \ref{sec:expansion}). One can use the transformation matrix $R$ to switch from one basis to another similar to Eq.
~({\ref{eq:GRAPE_gradient_originial_basis}}). Furthermore, Hessian-based optimization generally converges in fewer iterations than gradient-only, especially near the optimum where its convergence is quadratic. Last but not least, the higher accuracy of the calculation may help it altogether avoid local traps that can plague gradient-based quantum optimal control methods. 

Hence, we expect Hessian-based optimization to be an improvement over gradient-only unless the total number of steps $N$ is significantly large. We emphasize that for larger Hilbert-spaces, the exact matrix diagonalization via Eq. (\ref{eq:GRAPE_gradient}) becomes expensive, and other alternatives in the literature might be preferable \cite{goodwin2018advanced,jensen2020exact,floether2012robust} though the same general conclusions hold. 

In the following sections, we test and benchmark gradient-only vs. Hessian-based optimization on a standard quantum computational setup.

 \begin{figure*}
     \centering
     \includegraphics{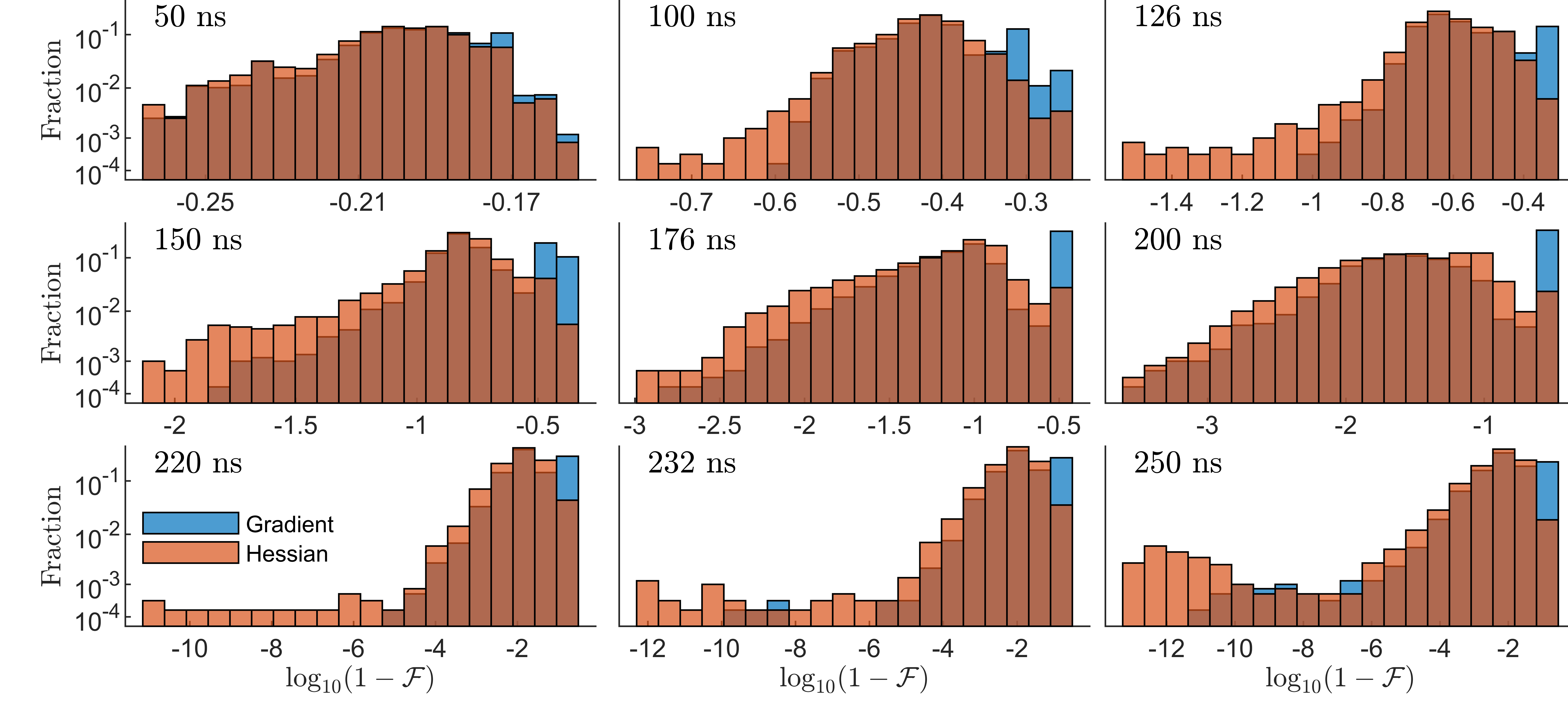}
     \caption{(Color online) Distribution of infidelities (lower is better) at different gate durations. Blue depicts gradient-only optimization, while orange depicts Hessian-based. Brown depicts the overlap of the two distributions. Gate durations are depicted in the figure.
     }
     \label{fig:histograms}
 \end{figure*}

\section{Transmon system} \label{sec:physical_system}

As a testbed for the Hessian-based optimization, we have chosen two transmon qubits dispersively coupled through a linear microwave resonator activated by a microwave field \cite{paraoanu2006microwave, rigetti2010fully, chow2011simple, goerz2017charting}. This type of setup is currently a frontier within superconducting circuit-based quantum information \cite{wendin2017quantum} and could enable the realization of many of the quantum technologies outlined in the introduction. Such a setup has previously been studied for gradient and machine learning based optimal control \cite{goerz2017charting,kirchhoff2018optimized,allen2019minimal,dalgaard2019global}.

Transmons are insensitive to charge noise, but suffer from having relatively low anharmonicity \cite{koch2007charge, theis2018counteracting}. We therefore include a third level for each transmon qutrit. We use an effective Hamiltonian, where we adiabatically eliminate the cavity and replace the qutrit-cavity coupling with an effective qutrit-qutrit coupling \cite{magesan2018effective}. Our starting point is to model each transmon as an anharmonic Duffing oscillator \cite{khani2009optimal} and describe the transmon-cavity coupling via the Jaynes-Cummings model, which in the absence of control is

\begin{equation}
    \begin{aligned}
        H_0 &= \sum_{j=1,2} \omega_j b_j^{\dagger}b_j + \frac{\delta_j}{2} b_j^{\dagger}b_j(b_j^{\dagger}b_j-1)
        + \omega_r a^{\dagger}a\\
        & + \sum_{j=1,2} g_j (a b_j^{\dagger} + a^{\dagger} b_j).
        \label{eq:actual_drift_Hamiltonian}
    \end{aligned}
\end{equation}
Here $b_j (b_j^{\dagger}) $ denotes the annihilation (creation) operator for the $j$th transmon in the $\{ \ket{00}, \ket{01},\ldots \ket{22} \}$ basis. We choose the transmon frequencies to be $\omega_1 /2\pi = 5.0 \si{\giga \hertz}$ and $\omega_2 /2\pi = 5.5 \si{\giga \hertz}$ with equal anharmonicities $\delta_1/2\pi = \delta_2/2\pi = -350 \si{\mega \hertz} $. For the cavity resonance frequency we choose $\omega_r/2\pi = 7.5 \si{\giga \hertz}$ with equal transmon-cavity couplings $g_1/2\pi=g_2/2\pi = 100  \si{\mega \hertz}$. These values are within typical experimental ranges (see e.g.~\cite{mckay2016universal}). 

In Appendix \ref{sec:circuit_qed_appendix} we derive an effective Hamiltonian where we eliminate the cavity and move into a rotating frame. We further drive the first transmon directly, which leads to the effective Hamiltonian

\begin{equation}
    \begin{aligned}
    H_{\text{eff}}(t) &= \Delta b_1^{\dagger} b_1 + \sum_{j=1,2} \frac{\delta_j}{2} b_j^{\dagger}b_j(b_j^{\dagger}b_j-1)\\
    & + J(b_1^{\dagger} b_2 + b_1 b_2^{\dagger}) + \Omega(t)( b_1^{\dagger} + b_1),
    \end{aligned}
    \label{eq:control_Hamiltonian}
\end{equation}
where $\Delta$ denotes the detuning between the transmons and $J$ denotes the effective coupling between the transmons (see Appendix \ref{sec:circuit_qed_appendix}). Here $\Omega$ denotes our control, which we limit to be in the range of $\Omega/2\pi = \pm 200 \si{\mega \hertz} $. This constrained amplitude, which will be enforced by the optimization, in part determines the minimal time, the quantum speed limit \cite{caneva2009optimal}, for which the control problem becomes exactly solvable. For many control tasks, optimal solutions contains segments that lie on the constraint-boundary, hence how the optimization algorithm handles constraints is of vast importance. In this work, we will attempt to make a V = \textsc{cnot} gate starting from the identity $U_0 = \mathbf{1}$ and so the projectors in Eq. (\ref{eq:Fidelity}) is with respect to the qubit subspace $\mathbb{P} = \sum_{i,j=0,1} \ket{i,j}\bra{i,j}$. We will steer the system via a single piecewise constant control, $ c_{j,1} = \Omega(t_j)$, as discussed in the previous section.

\section{Hessian vs. Gradient based optimization} \label{sec:Hessian vs. gradient}

We consider here the control problem outlined in the previous section, where we use piecewise constant pulses that consist of $\Delta t = \SI{2.0}{\nano \second} $ segments, which is within the bandwidth of standard microwave pulse generators. We compare the two approaches, gradient-only (i.e. \textsc{bfgs}) vs analytical Hessian-based optimization via an Interior-Point algorithm (see Appendix \ref{sec:different_optim}), over a wide range of different gate durations using multistarting. For each gate duration we use the two approaches to optimize the same 5000 seeds, which we draw uniformly at random. The relative optimality and step tolerance was set to $10^{-9}$ and $10^{-10}$ respectively (see documentation \cite{MATLAB}). We plot the results in Fig.~\ref{fig:grad_Hess_comparison_different_gate_times}(a)-(c), which respectively shows the best infidelity over the entire range of gate durations, the mean infidelity per seed, and the mean wall time consumption per seed for each gate duration. The two approaches often find different solutions in the optimization landscape, despite starting from the same initial seed. This stresses the need for selecting the optimization algorithm with care, and also motivates the comparison given in Appendix \ref{sec:different_optim}. 

Fig.~\ref{fig:grad_Hess_comparison_different_gate_times}(a), we see that when the gate duration increases above $\SI{200}{\nano \second}$ the control problem becomes exactly solvable in the sense that the best infidelities become insignificantly low. In the literature the minimal amount of time for which the problem becomes exactly solvable is termed the quantum speed limit \cite{caneva2009optimal,hegerfeldt2013driving,goerz2017charting, sorensen2018quantum}. Here we see that the Hessian-based optimization is able to come closer to the (unknown) quantum speed limit relative to gradient-only. Even in the low infidelity regime (i.e. below $10^{-8}$), Hessian-based optimization reaches a few orders of magnitude improvement over gradient-only. 

Fig.~\ref{fig:grad_Hess_comparison_different_gate_times}(b) depicts the mean infidelity for the two algorithms. The two algorithms perform equally well on average at shorter gate durations. In contrast, Hessian-based optimization performs better at longer gate durations, with mean infidelities being 3-4 times lower than gradient-only optimization. 

Fig.~\ref{fig:grad_Hess_comparison_different_gate_times}(c) depicts the mean wall time consumption for each gate duration. As already elaborated on, the computationally cost of the gradient scales linearly $\mathcal{O}(N)$ while the Hessian scales quadratically $\mathcal{O}(N^2)$. However, Hessian-based optimization generally requires fewer iterations for convergence than \textsc{bfgs}. For this reason, Hessian-based optimization actually turns out to be comparable with gradient-only at very short gate durations, where the Hessian is cheaper to calculate but the number of iterartions needed remains significantly lower. For longer gate durations, however, Hessian-based optimization becomes 2-3 times slower than gradient-only. 

In summary, Fig.~\ref{fig:grad_Hess_comparison_different_gate_times} shows that there are two distinct regimes, where the analytical Hessian of the unitary dynamics is preferable to its \textsc{bfgs}-approximation, but for different reasons. For fewer than $\SI{50}{\nano \second}$ (or 25 control steps), the two approaches performs equally well with respect to infidelity, however, the figure and scaling arguments indicates a potential speed advantage for the Hessian-based approach. Above this time, and especially near the quantum speed limit, we see the exact Hessian will produce better gate infidelities, both on average and for the best case although at the price of being slower. 

We plot in Fig.~\ref{fig:histograms} histograms over the infidelity distributions for the same data presented in Fig.~\ref{fig:grad_Hess_comparison_different_gate_times}. At $\SI{50}{\nano \second}$ in Fig.~\ref{fig:histograms}(a) we see two almost identical distributions. In contrast, at larger gate durations we see that Hessian-based optimization generally performs better than gradient-only both in terms of the quality and quantity of the best solutions found.

From Figs.~\ref{fig:grad_Hess_comparison_different_gate_times} and \ref{fig:histograms} we see that the two approaches converge to different minima although starting from the same initial seed. We also see that the Hessian-based solutions tend to be better i.e. reach lower infidelity compared to gradient-only, which is clear from the bimodal distributions in Fig.~\ref{fig:histograms}. We attribute this to the fact that Hessian-based optimization obtains more information about the curvature of the optimization landscape through the second derivative relative to gradient-only, which only has approximate knowledge of the second derivatives. This can lead gradient-only to slow down, or even be unable to converge at all. It may also cause trapping in local suboptimal solution as illustrated in Fig.~\ref{fig:pulse_ex_and_grad_v_Hess}(b).  

\begin{figure}
    \centering
    \includegraphics{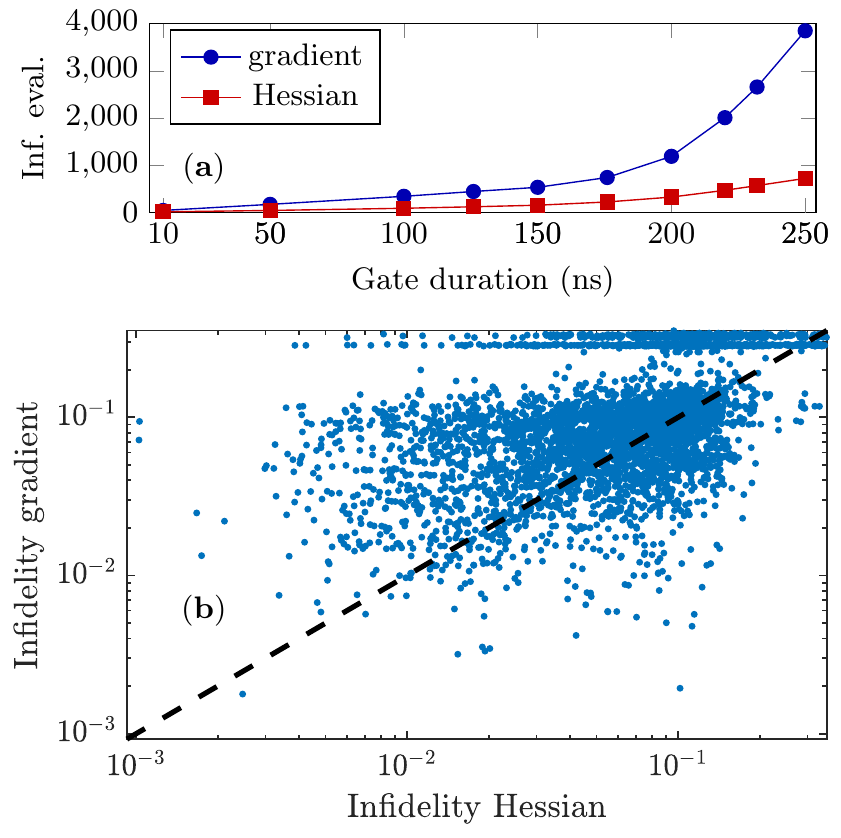}
    \caption{$\mathbf{(a)} $ The average number of infidelity evaluations for gradient-only and Hessian-based optimization. The figure verifies that gradient-only scales worse than Hessian-based with respect to the number of infidelity evaluations. 
    $\mathbf{(b)}$
    A scatter plot of the optimized infidelities from Fig.~\ref{fig:histograms} at $\SI{176}{\nano \second}$. The x-axis is the Hessian-based optimized solutions while the y-axis is for the gradient-only method. Solutions that lie on the dashed diagonal line corresponds to gradient and Hessian-based doing equally well. Above the diagonal is where the Hessian did best and vice versa.}
    \label{fig:fun_eval}
\end{figure}

In Fig.~\ref{fig:fun_eval}(a), we also depict the average number of infidelity evaluation used by either approach. Here we clearly see that the Hessian-based optimization uses fewer infidelity evaluations in comparison to the gradient-only. This illustrates the difference between quadratic convergence (Hessian) vs. superlinear convergence (gradient) as elaborated on in section \ref{sec:optimal_control_of_unitary}. 

To verify that the two approaches indeed do find different solutions, we plot in Fig.~\ref{fig:fun_eval}(b) a scatter plot of the optimized infidelities for the two approaches at $\SI{176}{\nano \second}$. If a given point lies on the dashed diagonal it would imply that the two approaches (most likely) found the same solution, as the figure shows this is most often not the case.  

Although the Hessian is more expensive to calculate than the gradient, we showed that Hessian-based optimization can be advantageous over gradient-only for even over 100 optimization parameters, which allows for reasonably complicated but still practical pulses that may easily be generated via modern arbitrary waveform generators. However, for a much larger number of parameters, Hessian-based optimization may become too slow to use at one point, even if the final result would be better. Hence to include the Hessian is a question of balancing the quantity and quality of found solutions.   
When using only a relative few seeds (e.g. due to a large amount of optimization parameters or complicated state spaces that require mores exploration), it may be very unlikely to reach near the global optimum for a practical wall time: In this case a gradient-only approach may be expected to produce faster if less reliable results. 

Even so, many optimization tasks within quantum control have less than 125 optimization parameters. For some control tasks, as the one considered here, increasing this number would make experimental implementation and calibration infeasible. Hence, we believe that explicitly incorporating the analytical Hessian of the unitary dynamics is advantageous for many quantum control tasks.

Finally, one may link the apparent utility of the Hessian to the nature of the control landscape. Indeed, the smoothness of the landscape \cite{hardy2001quantum} has earlier been argued to provide a computational advantage for quantum optimization \cite{rabitz2004quantum,caneva2009optimal,caneva2011chopped}. 
In the present work, we have explored the case where the cost functional is strongly constrained. Although such constraints formally remove the possibility of a global convergence \cite{zahedinejad2014evolutionary,Zhdanov2015wrong}, in our context the functional smoothness can nonetheless provide a mechanism to greatly speed up convergence alongside a multistarting strategy with analog controls \cite{zhdanov2017structure}.

\section{Conclusion}

In this paper, we have obtained the Hessian of the unitary dynamics as an extension to the widely used gradient-based, quantum optimization algorithm \textsc{grape}. Our calculations, which were based on diagonalization, revealed that the Hessian may be computed efficiently, with a high level of re-usability of already-calculated quantities obtained when evaluating the gradient. We believe our efficient calculation is advantageous to previous proposals and it allowed us to demonstrate improvements over gradient-only optimization in equal wall-time simulations without any code parallelization. We optimized a \textsc{cnot} gate on a circuit \textsc{qed} system consisting of two coupled transmon qubits. Here we demonstrated that a fast \textsc{cnot} gate is in principle feasible using only a single control on one qubit driven cross-resonantly.

For the numerical optimization, we used an Interior-Point algorithm, with either an analytically-exact Hessian or the Hessian-approximation scheme \textsc{bfgs} that only relies on the gradient. We compared the two approaches, Hessian-based or gradient-only, for a wide range of different gate durations. Since the Hessian contains squared the number of elements of the gradient, it is more expensive to calculate per iteration. However, Hessian-based optimization generally uses fewer iterations to converge. Moreover, the convergence occurs with greater accuracy, which can improve the quality as well as the quantity of good solutions.

We have found that, depending on the number of controls, either the wall time or fidelity of the solutions can be improved compared to the gradient. This appears to be generally true, although for very complex spaces where multistarting is not appropriate the gradient may be the only practical choice.  Nonetheless, for over 100 controls, we were still able to collect statistics over many seeds. We found that below 25 controls, the Hessian enabled faster convergence towards extrema. For more controls (and for the experimentally most interesting regime near the quantum speed limit), incorporating the Hessian provided a higher percentage of good solutions, often accompanied by bimodal distributions pointing to avoiding local trapping. Thus, best-case error was also seen to be improved.

\section{Acknowledgements}
The authors would like to thank Jens Jakob S\o rensen for participating in constructive scientific discussions. This work was funded by the ERC, H2020 grant 639560 (MECTRL), the John Templeton, and Carlsberg Foundations. The numerical results presented in this work were obtained at the Centre for Scientific Computing, Aarhus phys.au.dk/forskning/cscaa.

\newpage
\onecolumngrid
\appendix
\section{State transfer} \label{sec:state_to_state}
When only multiplication is needed in the propagation (as for typical expansions, see Appendix C), then this can be a preferred method to use for propagation of only one or a few orthogonal states (essentially specified by the projector $\mathbb{P}$ in \eqref{eq:Fidelity}). More concretely, we can rewrite
\begin{equation}
    \mathcal{F} =  \Big|\frac{1}{\text{dim}}\sum_{k=1}^{\text{dim}}  \bra{\psi_k} V^{\dagger}U\ket{\psi_k}\Big|^2,
    \label{eq:stateFidelity}
\end{equation}
In this case, the propagator scales as $O(n^2)$, $n$ the linear dimension, rather than exponentiation and matrix-matrix multiplication which typically scale as $O(n^3)$ (or $O(n^{2.8})$) for dense matrices. Our method is easily adaptable by replacing the left $U^L_j$ and right $U^R_j$ unitaries by left $\bra{\psi^{L}_j}=\bra{\psi_{targ}}U^L_j=\bra{\psi_{0}}V^{\dagger}U^L_j $ and right $\ket{\psi^{R}_j}=U^R_j\ket{\psi_{0}}$ states, respectively. $\bra{\psi^{L}_j}$ and $\ket{\psi^{R}_j}$ are dynamically calculated, as were $U^L_j$ and $U^R_j$, using a recursive approach. Similarly, the second derivatives can recursively use the time-propagated first-order derivatives, as needed for \eqref{eq:second_derive_U_different_steps}, as in
\begin{align} 
     \frac{\partial^2 \ket{\psi^{R}_j}}{\partial c_{i,k'} \partial c_{j,k} } 
    =\frac{\partial U_j}{\partial c_{j,k}} U_{j-1}\dots U_{i+1}
    \frac{\partial U_i}{\partial c_{i,k'}} \ket{\psi^{R}_{i-1}} = \frac{\partial U_j}{\partial c_{j,k}}\ket{\frac{\partial \psi^{R}_{j-1}}{\partial c_{i,k'}}}
\label{eq:states_second_derive_U_different_steps}
\end{align}
which uses only matrix-vector multiplication. Thanks to the recursive definition $\ket{\frac{\partial \psi^{R}_{j+1}}{\partial c_{i,k'}}}=U_{j+1}\ket{\frac{\partial \psi^{R}_{j}}{\partial c_{i,k'}}}$, this ends up once again
reducing the algorithm complexity with respect to the number of time steps (as in the main text) from $O(N^3)$ to $O(N^2)$. Using state propagation thus permits avoiding  matrix-matrix multiplication.
See also the original Khaneja \emph{et al.} GRAPE paper \cite{khaneja2005optimal}.

\section{Hessian calculation using eigenbasis} \label{sec:Hess}
The starting point of these calculations is Eq. (\ref{eq:double_integrals}), which can be written as

\begin{align}
    \Braket{m|
    \frac{\partial^2 U_j}{\partial c_{j,k'} c_{j,k}}
    |n}
    = \sum_{n'} \Big(
    H_{k'}^{(m,n')} H_{k}^{(n',n)} \mathcal{I}(n,n',m)
    +
    H_{k}^{(m,n')} H_{k'}^{(n',n)} \mathcal{I}(m,n,n')
    \Big).
    \label{eq:double_integral_short_form}
\end{align}
Here we have defined

\begin{align}
    \mathcal{I}(n,n',m) = (-i \Delta t)^2
    e^{\lambda_n}
    \int_0^1 d\alpha \alpha e^{\alpha(\lambda_{n'}-\lambda_n)}
    \int_0^1 d\beta
    e^{\beta \alpha(\lambda_m - \lambda_{n'})}.
\end{align}
In order to evaluate the above we must consider the the five different scenarios: $m \neq n \neq n'$, $m=n=n'$, $m=n'\neq n$, $m\neq n = n'$, and $m = n \neq n'$. We start with the first one which is $m \neq n \neq n'$. A direct calculation reveals

\begin{align}
    \mathcal{I}(n,n',m)
    = \frac{1}{E_m-E_{n'}}
    \Big[
    \frac{e^{-i E_{m} \Delta t} - e^{-i E_{n} \Delta t} }{E_m - E_n}
    -
    \frac{e^{-i E_{n'} \Delta t} - e^{-i E_{n} \Delta t} }{E_{n'} - E_n}
    \Big].
\end{align}
Note that the above expression is invariant under any permutation of the indices. Similarly, the second one, which is $m = n = n'$, gives

\begin{align}
    \mathcal{I}(n,n,n) = \frac{(-i \Delta t)^2}{2}
    e^{-i E_n \Delta t}.
\end{align}
The third one, which is $m = n' \neq n$, gives
\begin{align}
\mathcal{I}(n,m,m) = (-i \Delta t)^2
    e^{\lambda_n}
    \int_0^1 d\alpha \alpha e^{\alpha(\lambda_{m}-\lambda_n)}.
\end{align}
The antiderivative of $xe^{kx}$ is $\frac{(kx-1)}{k^2}e^{kx}$, which can be used to evaluate the above
\begin{align}
    \mathcal{I}(n,m,m) = 
    \frac{[-i \Delta t(E_m-E_n) - 1]e^{-i E_m \Delta t} + e^{-i E_n \Delta t}}
    {(E_m-E_n)^2}.
\end{align} 
The fourth one is $m \neq n' = n$
\begin{align}
    \mathcal{I}(n,n,m) 
    = 
    \frac{
    e^{-i E_m \Delta t} +
    (- i \Delta t
    [E_m-E_n]
    -1
    )
    e^{-i E_n \Delta t}
    }
    {(E_m-E_n)^2}.
\end{align}
And the last one, which is $m = n \neq n'$

\begin{align}
    \mathcal{I}(n,n',n)
    = 
    \frac{
    (-i\Delta t[E_n - E_{n'}] -1) e^{-i E_n \Delta t}
    + 
    e^{-i E_{n'} \Delta t}
    }
    {(E_n-E_{n'})^2}.
\end{align}
We can summarize the above results by writing

\begin{equation}
    \begin{aligned}
    \mathcal{I}(n_1, n_2, n_3)
    =
    \begin{cases}
    \frac{1}{E_{n_3}-E_{n_2}}
    \Big[
    I(n_3, n_2)
    -
    I(n_2, n_1)
    \Big],
    
    &\text{ if } n_1 \neq n_2 \neq n_3
    \\
    \\
    \frac{-i \Delta t}{2}
    I(n_1,n_1), &\text{ if } n_1 = n_2 = n_3
    \\
    \\
    \frac{1}{E_n - E_m}
    \Bigg[
    I(n,m) - I(m,m)
    \Big],
    &
    \begin{cases}
     \text{ if } n = n_1 \neq n_2 = n_3 = m\\
     \text{ or if } n = n_2 \neq n_1 = n_3 = m\\
     \text{ or if } n = n_3 \neq n_1 = n_2 = m.
    \end{cases}
    \end{cases}
\end{aligned}
\label{eq:hessian_integral}
\end{equation}
We may evaluate the above integral efficiently by storing the already calculated integrals $I(m,n)$ from Eq. (\ref{eq:Integral_gradient}). Also note that above integral is independent of the order of coefficients $n_1, n_2, n_3$. This implies that we may calculate the above efficiently and in advance of evaluating the matrix elements of the second derivative. Since $\mathcal{I}(n_1,n_2,n_3)$ is independent of the order of the indices, we may also take the integral out of a parenthesis in Eq. (\ref{eq:double_integral_short_form}) and hence we obtain Eq. (\ref{eq:U_final_expression_of_elements}).

\section{Hessian calculation using (Taylor) expansion} \label{sec:expansion}
The analytical gradient and Hessian of the propagation are intricately linked to how the evolution is calculated. In the main text we used eigenbasis decomposition, which is exact to numerical precision. In certain cases with e.g. large and/or sparse Hamiltonians approximate expansions such as Taylor, BCH, Pade, Suzuki-Trotter, or Chebychev may be preferable (especially in combination with matrix-vector algebra, Appendix A) \cite{ndong2009chebychev, goodwin2018advanced, jensen2020exact}.

As a supplementary example to the diagonalization-based approach, we consider here also the case using Taylor expansion of the propagation \cite{goodwin2018advanced}. Let such an expansion $\tilde U_j$ be given by (with $c_{j,0}=1$)
\begin{eqnarray}
     U_j \approx \tilde U_j &&= \sum_{l=0}^{L}\frac{1}{l!}(-i H(t_j) \Delta t)^l= \sum_{l=0}^{L}\frac{1}{l!}(\sum_{k=0}^M -i c_{j,k} H_k \Delta t)^l\\
        &&= \sum_{l=0}^{L}\frac{1}{l!}\sum_{l_0+l_1+\dots + l_M=l}\frac{l!}{l_0!l_1!\dots l_M!}\prod_{k=0}^{M}(-ic_{j,k} H_k \Delta t)^{l_k}
\end{eqnarray}
where we have used the multinomial theorem for the last equality. Here $L$ denotes a truncation parameter. The gradients and Hessians are calculated exactly as before in the main text, i.e.~Eqs.~\eqref{eq:second_derive_U_different_steps} and \eqref{eq:second_derive_U_same_steps}, where only the first and second derivatives of the single-step unitaries need to be calculated differently. Then,
\begin{eqnarray}
     \frac{\partial^r \tilde U_j}{ (\partial c_{j,k'})^r}=\sum_{l=0}^{L}\frac{1}{l!}\sum_{l_1+l_2+\dots + l_M=l}\frac{l!}{l_1!l_2!\dots l_M!}\frac{l_{k'}(l_{k'}-1)^{r-1}}{(c_{j,k'})^r}\left[\prod_{k=0}^{M}(-ic_{j,k} H_k \Delta t)^{l_k}\right]
\end{eqnarray}
where $r=1$ or $r=2$. Meanwhile
\begin{eqnarray}
     \frac{\partial^2 \tilde U_j}{\partial c_{j,k'} \partial c_{j,k''}}=\sum_{l=0}^{L}\frac{1}{l!}\sum_{l_1+l_2+\dots + l_M=l}\frac{l!}{l_1!l_2!\dots l_M!}\frac{l_{k'}l_{k''}}{c_{j,k'}c_{j,k''}}\left[\prod_{k=0}^{M}(-ic_{j,k} H_k \Delta t)^{l_k}\right]
\end{eqnarray}
when $k''\neq k'$.

Any such expansion for the unitary evolution has its own exact analytical formulas for gradient and Hessian of the evolutions w.r.t.~the controls. Note also that even for a very complex method one can always use automatic differentiation as a substitute, which is also exact as above, though we believe this is not strictly necessary or faster in any particular case. Thus in both cases, regardless of how the Hamiltonian exponential is calculated, one can still benefit from the derivatives \eqref{eq:second_derive_U_different_steps} and \eqref{eq:second_derive_U_same_steps}.

Note that, while the analytical gradient and Hessian are exact, the expansions themselves are not, which means fixed expansions must be used if monotonic convergence is to be guaranteed (i.e.~keeping $L$ fixed above), or else machine-precision-level error tolerance must be enforced. Otherwise, the directions of the gradient and Hessian may change when e.g.~an extra term is added to the expansion or a term is modified (as would be needed to satisfy adaptive error tolerance criteria for the expansion).

\section{Benchmarking different optimization algorithms} \label{sec:different_optim}
In the main text we considered synthesizing a \textsc{cnot} gate using piecewise constant pulses with $\Delta t = \SI{2.0}{\nano \second}$ via the \textsc{grape} algorithm. \textsc{grape} relies on a numerical optimization algorithm at its backend, for instance in the main text we used Interior-Point \cite{byrd1999interior, waltz2006interior} implemented in \textsc{matlab}'s library \textit{fmincon} \cite{MATLAB}. Interior-Point can either be supplied with the gradient and the Hessian-approximation scheme \textsc{bfgs} or the exact Hessian. 

To justify this specific choice we benchmark Interior-Point with other conventional optimization algorithms. Another choice, also implemented in MATLAB’s \textit{fmincon} \cite{MATLAB} for constrained optimization, is the Trust-Region-Reflective algorithm \cite{coleman1994convergence,coleman1996interior}. Similar to Interior-Point, Trust-Region-Reflective can be supplied with either the gradient and optionally the Hessian. We also benchmark against an unconstrained optimization algorithm quasi-Newton \cite{fletcher2013practical} implemented in  MATLAB’s \textit{fminunc}, where we instead impose constraint-bounds by adding a penalty term to the cost-function. We choose a quadratic penalty function that is zero inside the admissible region and grows quadratically outside $J_{\text{penalty}} = \sigma (\Omega -\Omega_{\min / \max})^2$, where we set the penalty factor to $\sigma = 10^5$. 

\begin{figure}
    \centering
    \includegraphics{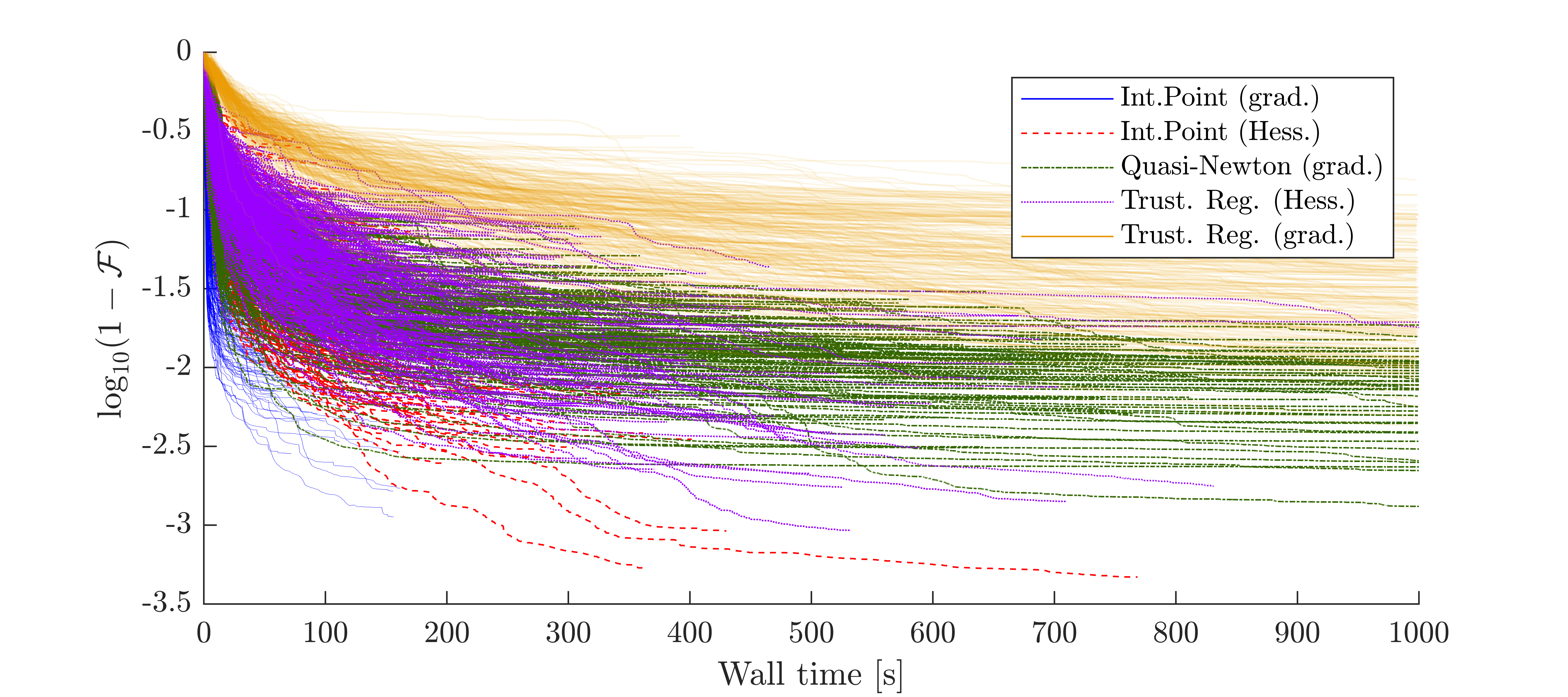}
    \caption{(Color online) A comparison between different optimization algorithms with and without the Hessian (see text).}
    \label{fig:diffent_opt_algs}
\end{figure}

For the comparison we consider the same control problem as in the main text with gate duration $T = \SI{200}{\nano \second}$. We let each optimization algorithm optimize the same 300 random seeds, which is drawn uniformly at random. The results is plotted in Fig.~\ref{fig:diffent_opt_algs} where we plot the infidelity for each seed as a function of wall time consumption, which we limit to $\SI{1000}{\second}$. From the figure we see that Interior-Point generally converges faster and at lower infidelity solutions. This justifies our choice of using Interior-Point for the results presented in the main text.

\section{Circuit \textsc{QED} calculations} \label{sec:circuit_qed_appendix}

The following derivation to some extent resembles the one given in Ref. \cite{magesan2018effective}. The starting point is the drift Hamiltonian given by Eq. (\ref{eq:actual_drift_Hamiltonian}). Here we eliminate the cavity by going to a frame rotating at $\omega_r$ via $R = \exp{[-i \omega_r (b_1^{\dagger} b_1 + b_2^{\dagger} b_2 + a^{\dagger} a )]}$. This gives

\begin{equation}
        H = \sum_{j=1,2} \Delta_j b_j^{\dagger}b_j + \frac{\delta_j}{2} b_j^{\dagger}b_j(b_j^{\dagger}b_j-1)
        + \sum_{j=1,2} g_j (a b_j^{\dagger} + a^{\dagger} b_j),
        \label{eq:drift_Hamiltonian_first_rotation}
\end{equation}
where $\Delta_j = \omega_j - \omega_r$. Then we perform a Schrieffer-Wolff transformation \cite{schrieffer1966relation} using $S = \sum_{j = 1,2} \frac{g_j}{\Delta_j} (a b_j^{\dagger} - a^{\dagger} b_j ) $ in order to eliminate the cavity. The resulting Hamiltonian, when any constant energy shifts have been removed and the cavity neglected, is

\begin{equation}
        H = \sum_{j=1,2} \Tilde{\omega}_j b_j^{\dagger}b_j + \frac{\delta_j}{2} b_j^{\dagger}b_j(b_j^{\dagger}b_j-1)
        + J (b_1^{\dagger} b_2 + b_1 b_2^{\dagger}).
        \label{eq:drift_Hamiltonian_SW}
\end{equation}
Here we see that the transmon-cavity coupling has been replaced with an effective transmon-transmon coupling where $J = \frac{g_1 g_2 (\Delta_1 + \Delta_2)}{\Delta_1 \Delta_2}$ and $\Tilde{\omega}_j = \omega_j + \frac{g_j^2}{\Delta_j}$ is now the dressed transmon state. The last step when transforming Eq. (\ref{eq:drift_Hamiltonian_SW}) into Eq. (\ref{eq:control_Hamiltonian}) is doing a second rotation $R' = \exp{[- i\Tilde{\omega}_2 (b_1^{\dagger} b_1 + b_2^{\dagger} b_2)]}$ such that the detuning becomes $\Delta = \Tilde{\omega}_2-\Tilde{\omega}_2$. We also add a direct drive on the first transmon $H_c(t) = \Omega(t) (b_1^{\dagger} + b_1)$ see e.g. Ref. \cite{magesan2018effective}. 

\section{Two-level example} \label{sec:two_control_example}
In Fig.~\ref{fig:pulse_ex_and_grad_v_Hess}(b), we illustrated three optimization methods based on gradient descent and Interior-Point with either \textsc{bfgs} or Hessian for a two-dimensional optimization problem. Here we have implemented a simple gradient descent algorithm using a fixed step size, where the step size is chosen to illustrate typical differences between gradient descent and \textsc{bfgs}. We briefly elaborate on what this Figure depicts. We consider a two-level Hamiltonian $H(t) = \sigma_x + c(t)\sigma_z$, with the goal of synthesizing a $X$ gate. We limit the control to two steps $N=2$, which reveals an analytical solution at $T_{\text{QSL}} = \pi / 2$ with $c_{1,1} = c_1 = 0$ and $ c_{2,1} = c_2 = 0$. At $T = 3 T_{\text{QSL}}$ the same solution is still optimal, but now several other solutions emergence in the optimization landscape, an effect also studied in Ref.~\cite{larocca2018quantum}. For instance the figure depicts a suboptimal solution at $(c_1, c_2) = (0.000, 2.285)$. We start the optimization near the two optima at $(c_1, c_2) = (-0.915,2.251)$ and plot the subsequent optimization. The two gradient-based optimization approaches fall into the nearest trap (i.e. suboptimal solution), while the Hessian optimization manages to avoid this solution. We attribute this result to the fact that the Hessian-based optimization has more information about the landscape curvature, which enables it to avoid the local trap and instead reach the optimal solution. The reader should of course keep in mind that this is but one example. 

\twocolumngrid
\bibliography{refs}

\end{document}